\documentclass[10pt, preprint]{aastex} 

\newcommand{\FeII}{Fe$\;${\small\rm II}\relax}

\newcommand{\CIII}{C$\;${\small\rm III}\relax}
\newcommand{\htwo}{H$_2$}
\newcommand{\HI}{H$\;${\small\rm I}\relax}

\newcommand{\DI}{D$\;${\small\rm I}\relax}
\newcommand{\OI}{O$\;${\small\rm I}\relax}

\newcommand{\NI}{N$\;${\small\rm I}\relax}
\newcommand{\lya}{Ly$\alpha$}
\newcommand{\lyb}{Ly$\beta$}
\newcommand{\lyg}{Ly$\gamma$}

\newcommand{\err}[2]{\ensuremath{^{+ #1}_{- #2}}}

\newcommand{\kms}{km~s$^{-1}$\relax}

\newcommand{\fuse}{{\em FUSE}\relax}
\newcommand{\iue}{{\em IUE}\relax}
\newcommand{\imaps}{{\em IMAPS}\relax}
\newcommand{\hst}{{\em HST}\relax}
\newcommand{\copernicus}{{\em Copernicus}\relax}

\begin{document}

\title{Deuterium and Oxygen Toward Feige 110:  Results from the Far Ultraviolet
Spectroscopic Explorer (\fuse) Mission}

\author{S.D.  Friedman\altaffilmark{1}, J.C.  Howk\altaffilmark{1}, P.
Chayer\altaffilmark{1,2}, T.M.  Tripp\altaffilmark{3}, G.
H\'ebrard\altaffilmark{4}, M.  Andr\'e\altaffilmark{1}, C.
Oliveira\altaffilmark{1}, E.B.  Jenkins\altaffilmark{3}, H.W.
Moos\altaffilmark{1}, W.R.  Oegerle\altaffilmark{5}, G.
Sonneborn\altaffilmark{5}, R.  Lamontagne\altaffilmark{6}, K.R.
Sembach\altaffilmark{1}, A.  Vidal-Madjar\altaffilmark{4}}

\altaffiltext{1}{Department of Physics \& Astronomy, 
The Johns Hopkins University, Baltimore, MD 21218; scott@pha.jhu.edu}

\altaffiltext{2}{Department of Physics \& Astronomy, University of Victoria,
P.O. Box 3055, Victoria, BC V8W 3P6, Canada}

\altaffiltext{3}{Princeton University Observatory, Princeton, NJ 08544}

\altaffiltext{4}{Institut d'Astrophysique de Paris, CNRS, 98 bis bld Arago,
F-75014 Paris, France}

\altaffiltext{5}{Laboratory for Astronomy and Solar Physics,
NASA/GSFC, Code 681, Greenbelt, MD 20771}

\altaffiltext{6}{Univ. de Montr\'eal, Dept. de Physique, CP 6128 Succ. 
Centre-Ville, Montr\'eal, PQ H3C 3J7 Canada}

\begin{abstract} 

We present measurements of the column densities of interstellar \DI\ and \OI\
made with the {\em Far Ultraviolet Spectroscopic Explorer} (\fuse), and of \HI\
made with the {\em International Ultraviolet Explorer} (\iue) toward the sdOB
star Feige 110 [{($l$,$b$) = (74\fdg09, -59\fdg07)}; d = $179 \err{265}{67}$ pc;
z = $-154 \err{57}{227}$ pc].  Our determination of the \DI\ column density made
use of curve of growth fitting and profile fitting analyses, while our \OI\
column density determination used only curve of growth techniques.  The \HI\
column density was estimated by fitting the damping wings of the interstellar
\lya\ profile.  We find log $N$(\DI) = $15.47\pm0.06$, log $N$(\OI) =
$16.73\pm0.10$, and log $N$(\HI) = 20.14$^{+0.13}_{-0.20}$ (all errors
2$\sigma$).  This implies D/H = $(2.14 \pm 0.82) \times 10^{-5}$, D/O = $(5.50
\err{1.64}{1.33}) \times 10^{-2}$, and O/H = $(3.89 \pm 1.67) \times 10^{-4}$.
Taken with the \fuse\ results reported in companion papers (Moos et al.  2001)
and previous measurements of the local interstellar medium, this suggests the
possibility of spatial variability in D/H for sight lines exceeding
$\sim$100 pc. This result may constrain models which characterize the mixing
time and length scales of material in the local interstellar medium.

\end{abstract}



\section{Introduction}

Precise measurements of primordial abundances of the light elements deuterium
(D), $^3$He, $^4$He, and $^7$Li relative to hydrogen have been a goal of
astronomers for many years.  In the standard Big Bang nucleosynthesis model,
these quantities are related in a straightforward way to the baryon-to-photon
ratio in the early universe, from which $\Omega_{\rm B}$, the fraction of the
critical density contributed by baryons, may be determined (Boesgard \& Steigman
1985).  In principle, a measurement of any single one of these abundance ratios
would be sufficient to determine $\Omega_{\rm B}$.  In practice, however, the
D/H ratio is more useful than the others for two reasons:  (1) D/H is a stronger
function of $\Omega_{\rm B}$ than the other light element ratios, and will yield
a more accurate value of $\Omega_{\rm B}$.  (2) The only appreciable source of
deuterium is the Big Bang itself.  No other significant production mechanisms
have been identified (Reeves et al.  1973; Epstein, Lattimer, \& Schramm 1976).
Since deuterium is easily destroyed in stellar interiors, the D/H ratio is
expected to monotonically decrease with time.  This makes the interpretation of
a D/H measurement simpler than the corresponding ratios for the other species.

Measurements of D/H in the local interstellar medium (ISM), as the present work
reports, have been made using \copernicus\ (Rogerson \& York 1973; see also the
review by Vidal-Madjar \& Gry 1984), \hst\ (e.g., Linsky et al.  1995;
Vidal-Madjar et al.  1998; Sahu et al.  1999), {\em ORFEUS} (G\"{o}lz et al.
1998; Bluhm et al.  1999), and \imaps\ (Jenkins et al.  1999; Sonneborn et al.
2000).  These local ISM measurements represent a lower limit to the primordial
value.  When corrected for the effects of astration (Tosi et al.  1998), the
results should be comparable to those obtained in low-metallicity, high redshift
\lya\ clouds (Kirkman et al.  2000; Burles \& Tytler 1998), which should be very
nearly the primordial value itself.

In this paper we present the results of an analysis of the sight line toward
Feige 110, an sdOB star located at a distance of $179 \err{265}{67}$ pc.  We
derive \DI\ and \OI\ column densities with data obtained from the {\em Far
Ultraviolet Spectroscopic Explorer} (\fuse), and estimate the \HI\ column
density with data from the {\em International Ultraviolet Explorer} (\iue).
This is the most distant target of the initial set of \fuse\ studies (H\'ebrard
et al.  2001; Kruk et al.  2001; Lehner et al.  2001; Lemoine et al.  2001;
Sonneborn et al.  2001; Wood et al.  2001) of deuterium in the local
interstellar medium.  The results of these studies are summarized by Moos et al.
(2001), who discuss the general nature and importance of the D/H problem, the
implications of the results in the context of previous D/H measurements, and how
they relate to other observational results, such the spatial variability of O/H
and N/H in the interstellar medium.

In \S 2 we present a summary of the observations and a description of the data
reduction processes.  In \S 3 the stellar model of Feige 110 is discussed, as
well as properties of the sight line and distance estimates to the star.  The
details of the analysis of the \DI\ and \OI\ column densities are given in \S4,
and the \HI\ analysis is presented in \S 5.  In \S6 we discuss the results of
this study.

\section{Observations and Data Processing}

The \fuse\ instrument has a bandpass of 905-1187 \AA.  It consists of four
co-aligned telescopes illuminating separate Rowland-circle spectrograph
channels.  Each channel is comprised of a diffraction grating and a portion of a
detector.  Each of the two large-area, two-dimensional detectors has two
microchannel-plate segments, designated A and B, which are coated with KBr
photocathodes.  The detectors record the position of each incident photon.  The
telescope mirrors and gratings of two channels are coated with SiC, and those of
the remaining two channels are coated with LiF over aluminum.  Detector 1
records the light from the SiC1 and LiF1 channels, and detector 2 from the SiC2
and LiF2 channels.  The dispersion of the SiC and LiF channels is 6.2 and 6.7
m\AA\ pixel$^{-1}$, respectively.  The \fuse\ resolution is approximately 15 -
20 \kms\ (FWHM).  A detailed description of the \fuse\ mission is given by Moos
et al.  (2000), and the instrumentation and performance is described by Sahnow
et al.  (2000).

Feige 110 was observed with \fuse\ for a total of 28 ksec under two separate
programs, M1080801 and P1044301.  The observation log is shown in Table 1.  All
data were obtained in histogram mode using the LWRS ($30 \arcsec \times 30
\arcsec$) aperture.

The data were reduced using CALFUSE pipeline version 1.8.7.  The individual
1-dimensional spectra were co-added to form the final spectrum for each channel
and detector segment separately after removing the relative shifts between
individual spectra on the detector caused by image and grating motion (Sahnow et
al.  2000).  The shifts were determined by cross-correlating the individual
spectra over a limited wavelength range which contained prominent spectral
features but no airglow lines.  Typical shifts were $\lesssim 4$ pixels, or
$\lesssim 0.025$\AA.  Although \fuse\ observations of \HI\ and \OI\ lines are
sometimes contaminated by airglow emission, requiring the use of data obtained
only during spacecraft orbital night, it was found that this was unimportant in
our analysis, and the combined day and night data were used.

Most of our analysis was based on SiC1B and SiC2A spectra ($910 \lesssim \lambda
\lesssim 1005$ \AA), which span the high order Lyman lines (Ly$\gamma$ through
the Lyman limit).  SiC2A generally has somewhat better spectral resolving power,
but also suffers from more detector fixed-pattern noise.  In rare cases,
fixed-pattern features can cause the detector response to vary by as much as
20\% over very small scales, and while the resulting spectral features are
sometimes diminished by the motion of the spectra on the detectors (Sahnow et
al.  2000) they are not explicitly removed in the current CALFUSE pipeline
reduction.  This difficulty is ameliorated by the multiple channel design of
\fuse; measurements of absorption lines that appear in the spectra of more than
one channel are independent, which can help to distinguish between real and
instrumentally-induced spectral features.

\begin{figure}
\epsscale{0.9}
\plotone{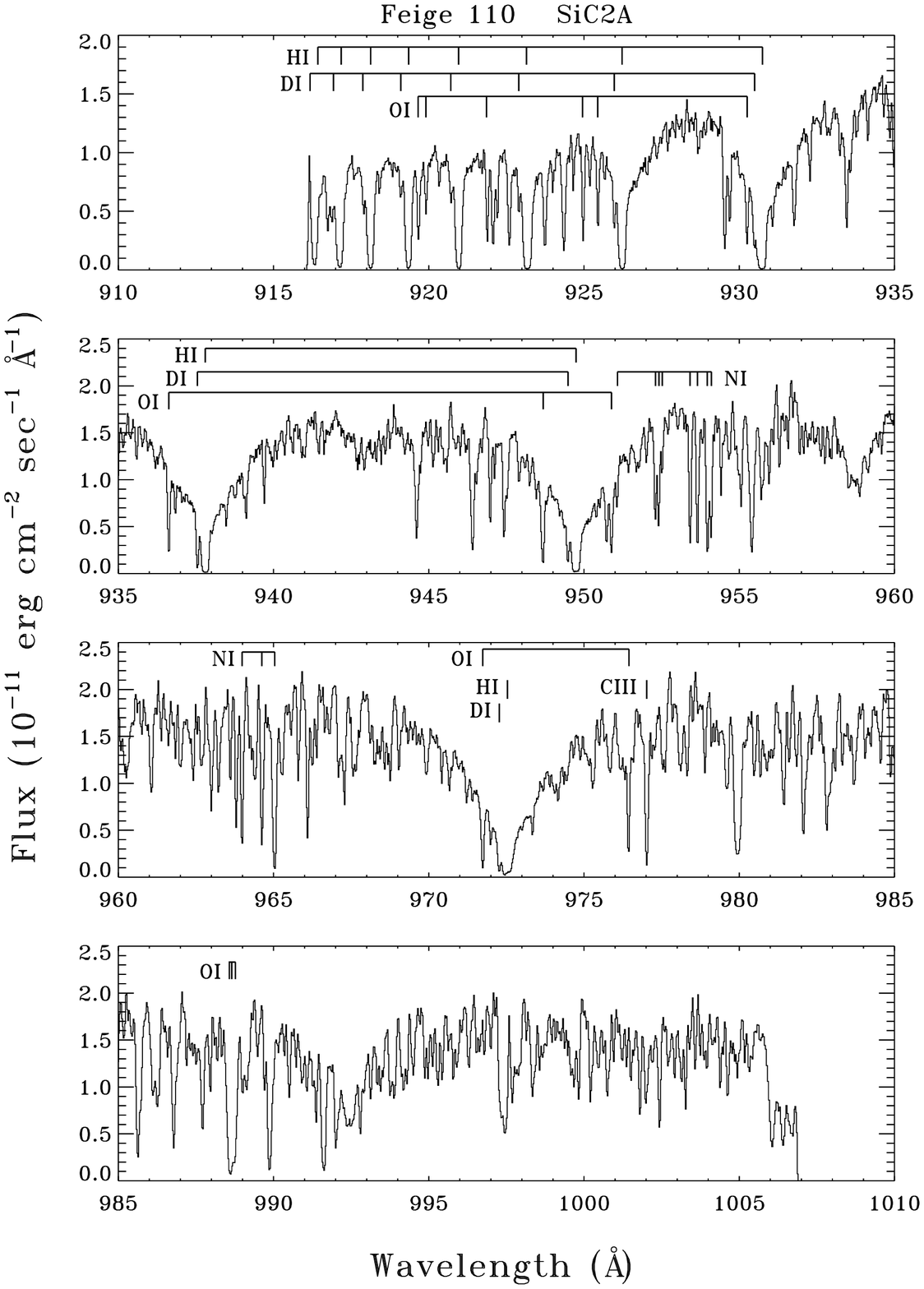}
\caption{The SiC2A spectrum, showing absorption lines used in the
analysis of the \DI\ and \OI\ column densities.  Interstellar \DI, \HI, \OI,
\NI, and \CIII\ lines have been marked.  Many of the remaining lines are due to
interstellar \htwo, or metal lines arising in the atmosphere of this subdwarf OB
star.  The data have been binned by 4 pixels ($\sim 0.024$ \AA) in this figure.
Note the flux scale change between the panels.
\label{fig_s2aspec}}
\end{figure}

Figure~\ref{fig_s2aspec} shows the co-added SiC2A spectrum of Feige 110, which
covers the wavelength range 916 - 1007 \AA.  \DI, \OI, and other lines arising
in the ISM are identified.  The spectral resolution is about 17 \kms (FWHM).
Typical S/N ratios for this observation are between 20 and 25.

The \HI\ column density along this line of sight was estimated using the Lyman
$\alpha$ profile from the single high resolution spectrum of Feige 110 taken
with \iue.  The 15.3 ksec observation was made on 15 October 1981 (see Table 1),
and reduced using IUESIPS.  The analysis of these data is discussed in more
detail in \S 5.

\section{The Line of Sight to and Stellar Spectrum of Feige~110}

Table 2 summarizes the important properties of Feige 110 and the sight line to
this star.  The trigonometric parallax of Feige~110 was measured to be $\pi =
5.59 \pm 3.34$ milliarcsec using the Hipparcos satellite (Perryman et al.\
1997), which yields a distance of $179 \err{265}{67}$ pc.  The star lies in the
direction ($l$,$b$) = (74\fdg09, -59\fdg07) at z = $-154 \err{57}{227}$ pc from
the Galactic plane, well beyond the local interstellar cloud (Lallement \&
Bertin 1992), and probably beyond most of the neutral hydrogen associated with
the wall of the Local Bubble (Sfeir et al.  1999).  This is not true of most of
the other local ISM sight lines studied with \fuse\ (see Moos et al.  2001 and
references therein).  Therefore, along this sight line we may be sampling
different material, depending on the mixing efficiency over these length scales
(Tenorio-Tagle 1996).  The average hydrogen density along this sight line is
$\langle n_H \rangle$ $\equiv$ $N$(\HI)/d $\approx$ 0.27 cm$^{-3}$, and Feige
110 is one of only two stars in the current \fuse\ deuterium sample to show
molecular hydrogen absorption (the other being BD+28\degr4211; see Sonneborn et
al.  2001).

Feige~110 was discovered by Feige (1958) in a search for underluminous hot stars
brighter than 14 mag.  In their study on the nature of faint blue stars,
Greenstein \& Sargent (1974) analyzed the optical spectrum of Feige~110 and
obtained a rough estimate of its atmospheric parameters.  Heber et al.\ (1984)
improved Greenstein and Sargent's analysis by using NLTE model atmospheres and
determined $T_{\rm{eff}} = 40$,000 K, $\log g = 5.0$, and $\log({\rm{He/H}}) =
-1.5$.  Based on the detection of strong Balmer lines, weak \ion{He}{1}
$\lambda$4471 and \ion{He}{2} $\lambda$4685 lines, Heber et al.\ classified
Feige 110 as a subdwarf OB star (sdOB).  The atmospheric parameters of Feige~110
put it at the hot end of the sdB/sdOB population, which includes He-burning
stars having masses close to 0.5 M$_\sun$ with very thin H-rich envelopes
($M_{\rm{env}} < 0.05$ M$_\sun$) (see, e.g., Caloi 1989; Dorman, Rood, \&
O'Connell 1993).

The parallax distance of Feige~110, 179 pc, is a factor of $\sim$5 smaller than
the distance computed using the atmospheric parameters obtained by Heber et al.\
(1984).  This discrepancy prompted us to reanalyze the optical spectrum of
Feige~110.  We used three optical spectra:  1) a {\it STIS} spectrum retrieved
from the Multimission Archive at the Space Telescope Science Institute
(MAST); 2) observations taken by one of us (R.L.)  at Cerro Tololo
Inter-American Observatory with the 4 m telescope; and 3) observations obtained
by R.\ Saffer (private communication) at Kitt Peak National Observatory's 2.1 m
telescope.

We estimated the atmospheric parameters of Feige~110 by comparing its optical
spectrum to a grid of synthetic NLTE stellar model atmosphere spectra.  We used
the programs TLUSTY/SYNSPEC (Hubeny \& Lanz 1995) to compute models having a
pure H and He composition.  The grid covered the temperature range 35,$000
{\rm{K}} \leqslant T_{\rm{eff}} \leqslant 50$,000 K in steps of 5,000 K for five
values of the surface gravity, $\log g = 4.5$, 5.0, 5.5, 6.0, 6.5, and for four
values of the helium abundance, $\log({\rm{He/H}}) = -1.0$, $-1.5$, $-2.0$, and
$-2.5$.  We obtained $T_{\rm{eff}} = 42$,$300 \pm 1,$000 K, $\log g = 5.95 \pm
0.15$, and $\log({\rm{He/H}}) = -1.95 \pm 0.15$, by using a $\chi^2$
minimization technique to fit the observed spectra to the synthetic spectra.
These parameters are the mean values obtained from the three optical spectra.
The larger gravity derived here implies a lower luminosity for Feige~110 than
that derived by Heber et al., and consequently a lower distance.  Our computed
distance ($d = 288 \pm 43$ pc), now agrees with the Hipparcos parallax distance
($d = 179 \err{265}{67}$ pc) within the stated uncertainties.

Feige 110 displays a very complex far-ultraviolet spectrum (Figure 1).  The
strong Lyman series lines of \HI\ (Ly$\beta$ up to Ly10) and the \ion{He}{2}
lines ($\lambda$1084 up to $\lambda$942) are the dominant stellar features.
We identified all the ISM lines superimposed on the stellar spectrum of
Feige~110, but were unable to identify many remaining photospheric lines.  This
identification is complicated by the lack of reliable atomic data, especially
for the iron-peak elements.  The strongest photospheric metal lines are the
\ion{N}{4} $\lambda$923 sextuplet, \ion{S}{6} $\lambda\lambda$933 and 944
doublet, \ion{N}{5} $\lambda$955, \ion{N}{3} $\lambda$979.9 quadruplet,
\ion{N}{3} $\lambda\lambda$989.80, 991.51, and 991.58 triplet, and \ion{P}{5}
$\lambda\lambda$1118 and 1128 doublet.  We identified photospheric lines from N,
S, Cr, Fe, and Ni.  Neither carbon (\ion{C}{3} $\lambda$1175) nor silicon
(\ion{Si}{4} $\lambda\lambda$1122.49, 1128.33, and 1128.49) was detected in the
\fuse\ spectrum.  This confirms the study of Heber et al.\ (1984) who reported
the non-detection of carbon and silicon in the \iue\ spectrum of Feige~110.

\begin{figure}
\plotone{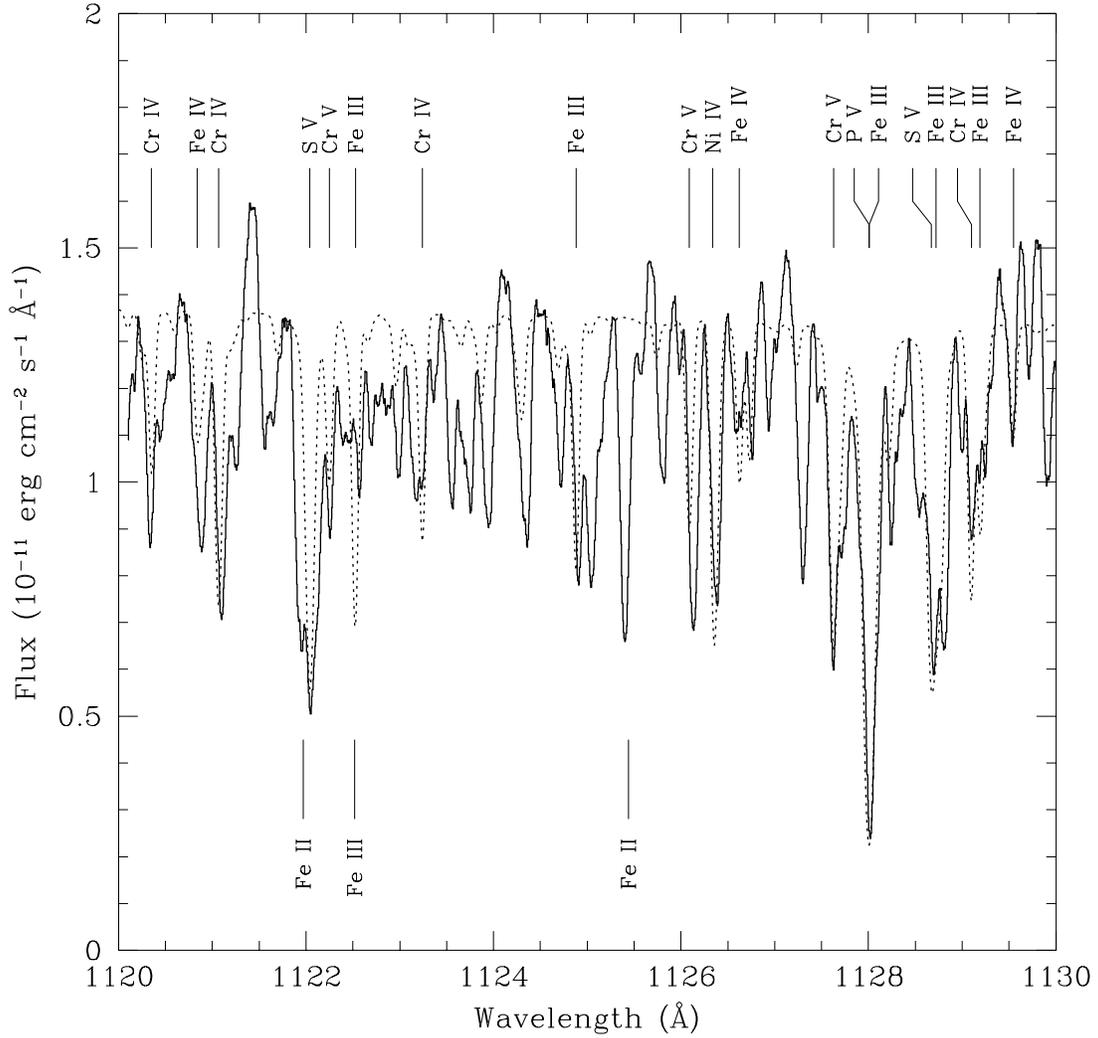}
\caption{Portion of the \fuse\ spectrum ({\it solid line}) compared to the
synthetic spectrum ({\it dotted line}).  The identified photospheric lines are
labelled above the spectrum, and the ISM lines are labelled below the spectrum.
The lack of reliable atomic data for the stellar lines is apparent in this
portion of the spectrum.  Note the absence of any \ion{Si}{4}
$\lambda\lambda$1122.49, 1128.33, and 1128.49 features in the \fuse\ spectrum,
as noted in the text.
\label{fig_spec_f110}}
\end{figure}

Figure~\ref{fig_spec_f110} shows a comparison between the synthetic spectrum and
a portion of the \fuse\ spectrum.  The synthetic spectrum is computed using the
atmospheric parameters listed in Table 2.  All the elements from H to Zn are
included with solar abundances, except for He (see Table 2), C ($\sim
4\times10^{-6}$ solar), Si ($\sim 2\times10^{-7}$ solar), and Cr (21 solar).
Figure~\ref{fig_spec_f110} illustrates the difficulty in matching the model with
the observed data, even though we use all atomic line data from the Kurucz \&
Bell (1995) database\footnote[1]{Atomic data from CD-ROM 23, which is available
at:  http://cfaku5.harvard.edu/cdroms.html}.  The non-identification of stellar
lines poses the problem of recognizing possible blends between the ISM and
photospheric lines.  Also, the placement of the continuum may be a problem in
regions of the spectrum where many stellar lines are not identified.  We take
these uncertainties into consideration in our determination of the interstellar
\DI\ and \OI\ column densities.

In addition to \HI, \DI, and \OI, detected interstellar lines include \htwo,
\NI, \ion{N}{2}, \FeII, \CIII, \ion{P}{2}, and \ion{Ar}{1}. However, reliable
determinations of the column densities of these species are difficult due to
saturation and blending effects, and have not been done for this study.

\section{$N$(\DI) and $N$(\OI) Toward Feige 110}

Absorption line studies of the interstellar medium are best done with background
sources consisting of hot stars with large $v$sin($i$) values or few visible
photospheric metal lines, both of which yield relatively smooth continua over
the scale of typical ISM absorption lines.  Unfortunately, Feige 110 does not
fit this description.  The far-ultraviolet continuum is very complex
(Figure~\ref{fig_s2aspec}), primarily due to the large number of metal lines
arising in the atmosphere of this sdOB star.  Analysis of the \DI\ and \OI\
column densities was possible only with a judicious choice of lines which were
not too severely blended with photospheric features.  Proper continuum placement
was somewhat uncertain and, in general, is the largest source of systematic
errors in our measurements.

Two techniques have been used to estimate the column densities in this analysis.
The first involves directly measuring the equivalent widths of absorption lines,
which are fit to a single-component Gaussian curve of growth (COG) (Spitzer
1978).  We are unaware of any high resolution optical or ultraviolet spectra
that could reveal the presence of multiple components along this sight line.
Furthermore, there is no evidence in the \fuse\ spectra of multiple components,
which would in any case be difficult to detect unless they were separated in
velocity by at least $\sim$10 \kms.  We therefore made the simplifying
assumption that a single interstellar cloud with a Maxwellian velocity
distribution is responsible for the absorption.  The estimates of the column
density ($N$) and Doppler parameter ($b$) derived from the COG are likely to
include systematic errors due to this assumption.  For example, the derived
$b-$value will not be a simple quadrature combination of thermal and turbulent
velocity components, as it would be for a true single cloud.  Instead, it will
also reflect the spread in velocities of the multiple clouds that are likely to
lie along a sight line of this length.  Jenkins (1986) has discussed the
systematic errors associated with the COG technique when there are multiple
clouds of various strengths along a sight line, and has shown that if many
non-saturated lines are used, and if the distribution of $N$ and $b-$values in
the clouds is not strongly bimodal, then the column density is likely to be
underestimated by $\la$ 15\%.

Despite these shortcomings, the COG technique has the virtue of being unaffected
in principle by convolution with the instrumental line spread function (LSF),
provided that such convolution does not lead to unacceptable blending with
neighboring lines.  However, if there is a very broad component to the LSF, then
in practice any technique may underestimate the column density of unsaturated
components because an erroneous continuum placement could mask some fraction of
the absorption.  This is particularly important since the \fuse\ LSF has not yet
been well characterized, though it is known to have both broad ($\sim$20 pixels)
and narrow ($\sim$10 pixels) components and to change as a function of
wavelength.

The second technique utilizes the profile fitting code {\tt Owens.f}, developed
by M.  Lemoine and the French \fuse\ Science Team.  This code models the
observed absorption lines with Voigt profiles using a $\chi^2$ minimization
procedure with many free parameters, including the line spread function, flux
zero point, gas temperature, and turbulent velocity within the cloud.  The
application of {\tt Owens.f} to \fuse\ data is described in greater detail by
Hebrard et al.  (2001).  Neither the COG nor profile fitting techniques include
oscillator strength uncertainties in the error estimates.

\subsection{The \DI\ Analysis}

\begin{figure}
\epsscale{0.95}
\plotone{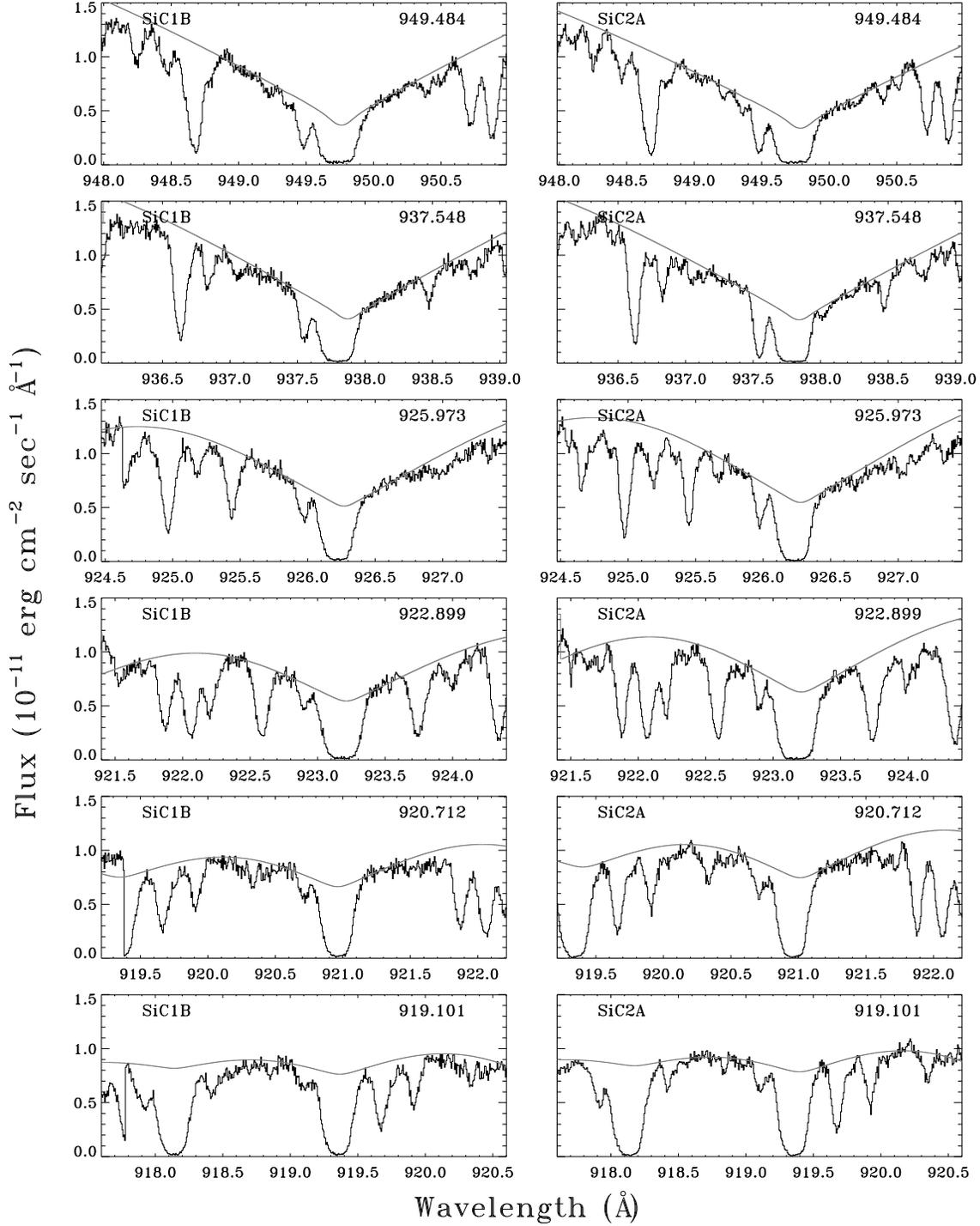}
\caption{The absorption line profiles used in the analysis of the \DI\ column
density.  The thin line in each panel shows the stellar model, which has been
shifted in wavelength and scaled in flux to match the continuum away from the
core of the interstellar \HI\ absorption.  The spectra from segment SiC1B are in
the left panels, and those from segment SiC2A are in the right panels.
\label{fig_stellar_cont}}
\end{figure}

To properly measure the \DI\ equivalent widths the effects of the stellar \HI\
Lyman series and \ion{He}{2} absorption must be removed.  This was done by
shifting in velocity the synthetic stellar model discussed in \S3 until the
stellar \HI\ damping wings matched the observed spectrum away from the cores of
the interstellar HI lines.  The resulting average velocity difference was
$\Delta V \equiv V_{\rm ISM} - V_* = -11\pm9$ \kms\ and $-15\pm4$ \kms\ for the
SiC1B and SiC2A spectra, respectively.  This compares well with $\Delta V =
-12$ \kms\ from measurements of five {Si$\;${\small\rm II}} ISM lines and two
{N$\;${\small\rm V}\relax} stellar lines in the \iue\ spectrum.
Figure~\ref{fig_stellar_cont} shows the \HI\ lines measured with \fuse\ as well
as the profiles from the stellar model.

An additional correction to the \DI\ absorption profiles was performed to remove
blends with adjacent molecular hydrogen absorption lines.  There are not enough
unblended \htwo\ lines to determine column densities of all the rotational
levels of \htwo.  However, only lines in the $J=2, 3, 4$ rotational levels of
\htwo\ are blended with the observed \DI\ lines, and the column densities are
small enough as to require only minor corrections.  Therefore, we fit Gaussians
to the observed profiles of unblended \htwo\ lines in the same rotational levels
as those overlapping the \DI\ transitions, and scaled by the relative $\lambda
f-$ values to estimate the strength of the contaminating \htwo\ lines.  The
estimated \htwo\ absorption was then divided out of the stellar-normalized \DI\
profile.  Table 3 lists the unblended \htwo\ lines and the measured equivalent
widths used to determine the \htwo\ corrections.

\begin{figure} 
\epsscale{0.85}
\plotone{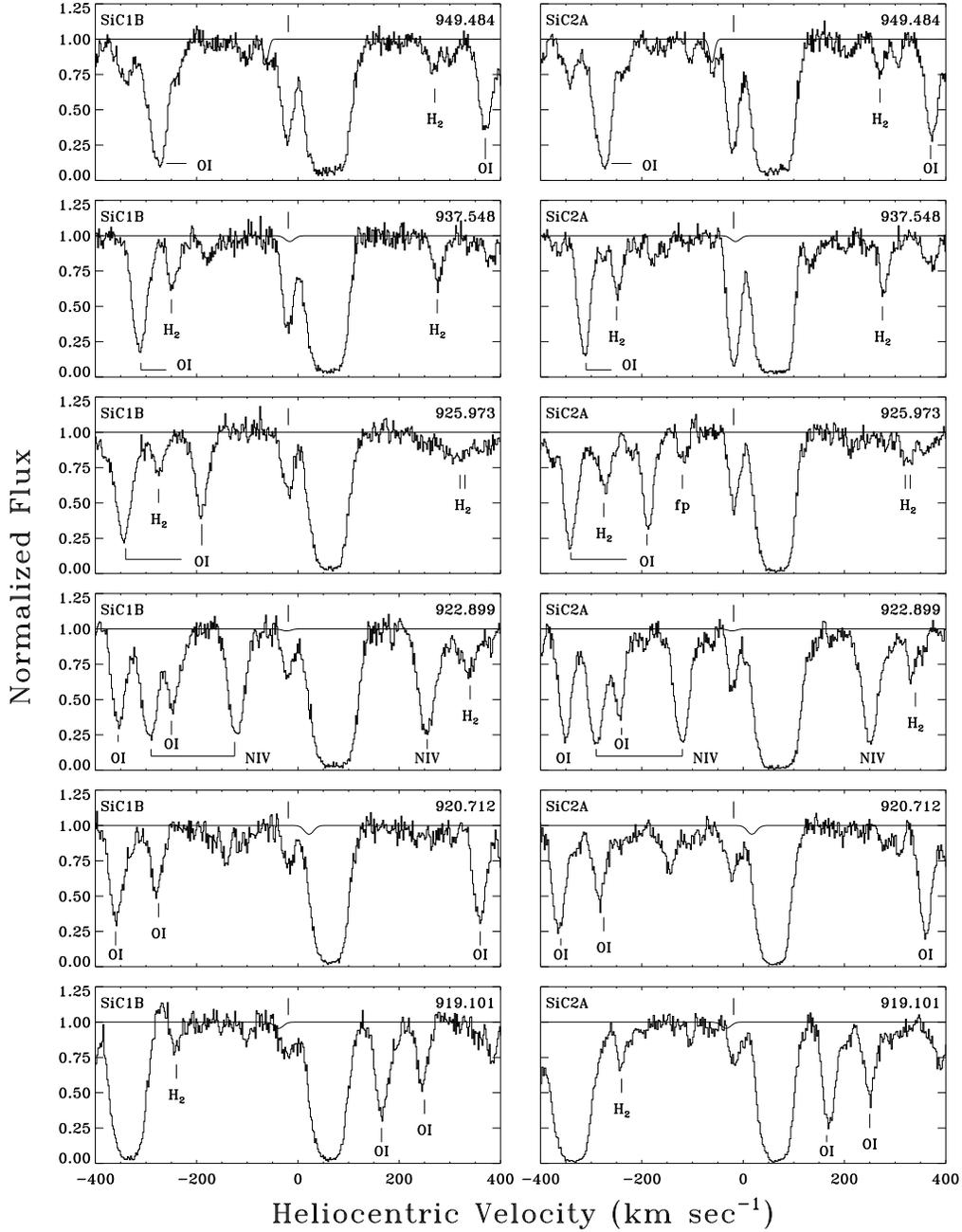}
\caption{Continuum normalized \DI\ absorption profiles, which are centered at a
velocity $V_{helio}=-19$ \kms.  The positions of the \DI\ lines are indicated by
vertical tick marks above the spectra.  Superimposed on the smooth continuum
line of each profile is the Gaussian \htwo\ line whose equivalent width and FWHM
have been calculated based on unblended \htwo\ lines in the same rotational
level, which are listed in Table 3.  (The \DI\ $\lambda$925.973 line is not
blended with \htwo.)  The $\lambda$937.548 line in the SiC2A spectrum is much
stronger than expected, and is probably contaminated by a detector defect.  It
has not been included in the \DI\ analysis.  A detector fixed-pattern feature is
denoted by "fp."  The spectra from segment SiC1B are in the left panels, and
those from segment SiC2A are in the right panels.
\label{fig_DI_cont}} 
\end{figure}

We measured the equivalent widths ($W_{\lambda}$) of the cleaned \DI\ lines
after fitting low-order Legendre polynomials to the local continuum profiles.
This process allowed us to remove small residual flux discrepancies between the
stellar model and the observed spectrum.  Figure~\ref{fig_DI_cont} shows the
continuum-normalized \DI\ profiles and our estimate of the \htwo\ contamination
of the \DI\ lines.  The interstellar lines are at a heliocentric
velocity\footnote{The local standard of rest velocity is $V_{LSR} =
V_{\sun}+2.6$ \kms.}  of $V_{\sun} = -19$ \kms, base on the \iue\ data.  Each
line was integrated over velocity limits that depended on blending and local
fixed-pattern noise.  The effects of scattered light are negligibly small for
the \DI\ COG analysis, and no correction was made.  The measured equivalent
widths are given in Table 4.  The estimated errors from this method (described
in detail by Sembach \& Savage 1992) include contributions from both statistical
and fixed-pattern noise in the local continuum, but do not include a
contribution from systematic continuum placement errors of the type considered
below.

\begin{figure}
\epsscale{0.90}
\plotone{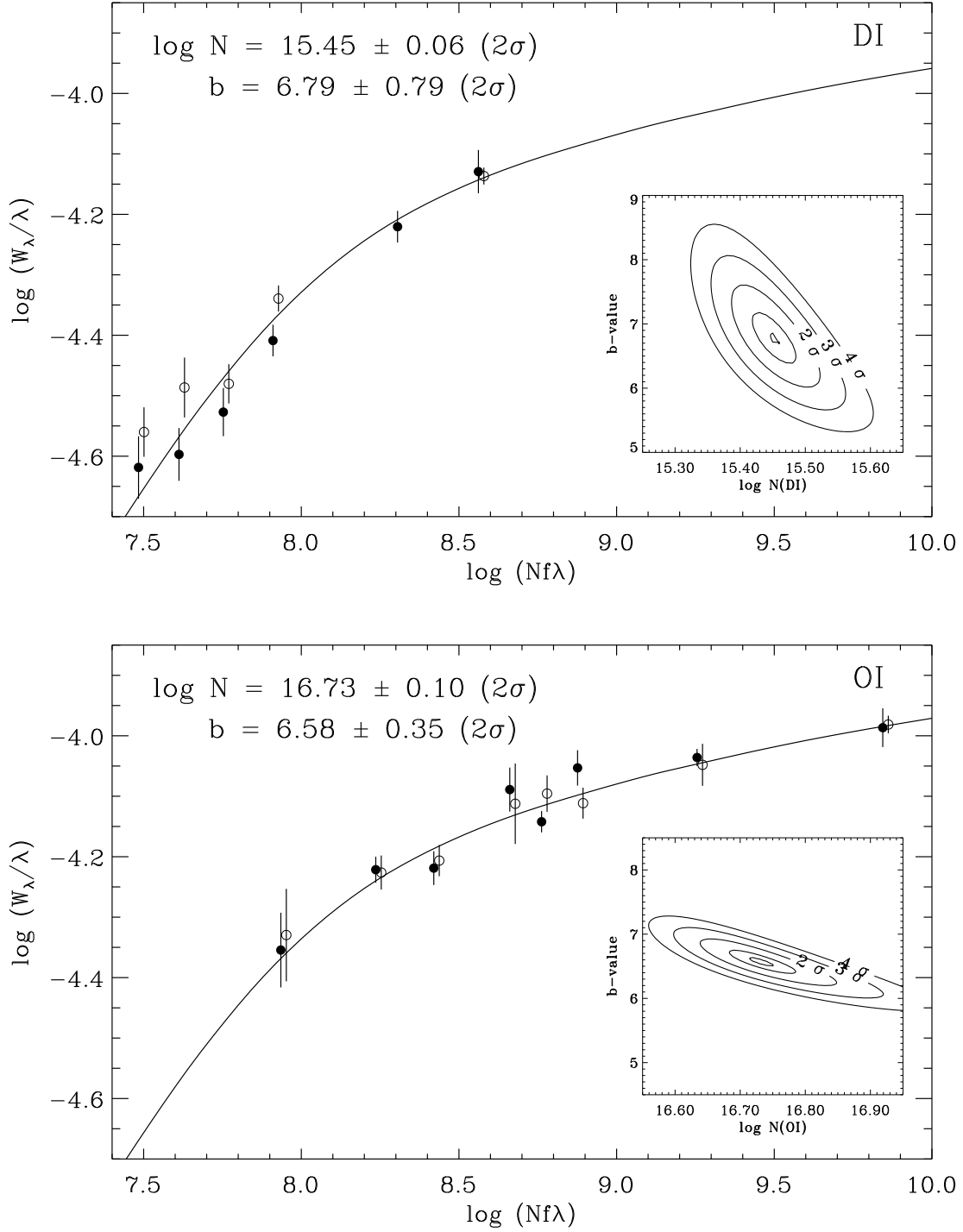}
\caption{The curves of growth for \DI\ (top) and \OI\ (bottom).  In both cases,
a single-component, Gaussian curve of growth has been assumed.  The insets show
the log($N$)/$b-$value error contours.  The large number of weak \DI\ lines
available tightly constrain the \DI\ column density.  Most of the \OI\ lines are
partially or fully saturated, so the column density is less well constrained.
For clarity, at each wavelength the data points derived from the SiC1B spectra
(solid circles) and the SiC2A spectra (open circles) have been slightly
separated horizontally.
\label{fig_cog}}
\end{figure}

The $W_{\lambda}$ measurements were placed on a best-fit curve of growth, as
shown in Figure~\ref{fig_cog}.  Lines from both the SiC1B and SiC2A spectra were
used, with the exception of the Ly$\epsilon$ 937.548 \AA\ line in SiC2A.  As
shown in Figure~\ref{fig_DI_cont} this line is much stronger than expected.  It
is stronger than the same line in the SiC1B spectrum, a comparison which
demonstrates the utility of the multi-channel design of \fuse, and stronger than
the Ly$\delta$ 949.484 \AA\ line, which cannot be correct.  We believe the
apparent strength of the line is due to a fixed-pattern effect on the detector,
and this line is ignored in the remaining analysis.

The systematic error on each equivalent width measurement is dominated by
uncertainties in the placement of the continuum.  We estimate the systematic
continuum placement uncertainties using the following procedure.  First, the
best estimate of $W_{\lambda}$ for each line was determined using what we
consider to be the proper continuum placement, and integrating over an
appropriate velocity interval.  Next, the maximum plausible $W_{\lambda}$ was
determined by placing the continuum at its maximum reasonable position in the
region of the absorption line, and integrating over the largest reasonable
velocity interval.  Then the minimum plausible $W_{\lambda}$ was measured using
a similar procedure.  The difference between these maximum and minimum values
and the best value, whichever is greater, is taken to be a $2\sigma$ systematic
error.  Half of this value is combined in quadrature with the statistical error
on the equivalent width measurement to give the total $1\sigma$ uncertainty.
The column density computed using the COG analysis is log
$N$(\DI)$_{COG}$=$15.45\pm0.06$ ($2\sigma$) and the Doppler parameter is
$b$=$6.79\pm0.79$ \kms ($2\sigma$).

\begin{figure} 
\plotone{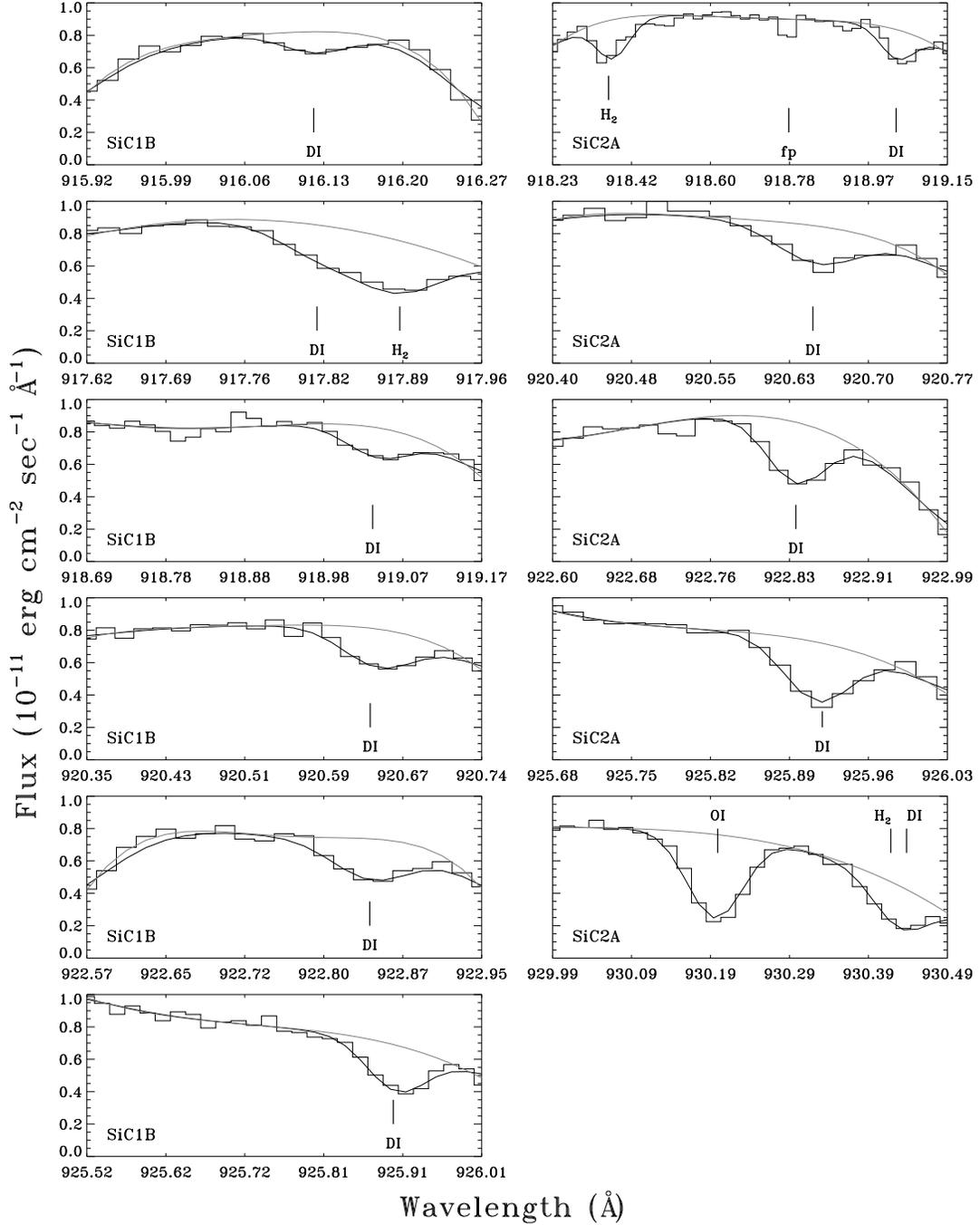}
\caption{Fits to the \DI\ lines using the {\tt Owens.f} profile fitting code.
The histogram line is the observed spectrum and the thin black line is the
computed fit.  The light gray line shows the continuum.  A detector
fixed-pattern feature is denoted by "fp."  The spectra from segment SiC1B are in
the left panels, and those from segment SiC2A are in the right panels.
\label{fig_owens}} 
\end{figure}

An independent calculation of the \DI\ column density was made using the profile
code {\tt Owens.f}.  Seven \DI\ lines in eleven separate spectral windows from
the SiC1B and SiC2A spectra were simultaneously fit to the stellar continuum
normalized profiles described above.  Each line was constrained to a common
value of the radial velocity, column density, and temperature.  The continuum
shape, zero flux level, and LSF were allowed to vary in each spectral window
encompassing a \DI\ line.  The LSF was modelled as a Gaussian, with a FWHM that
was in the range $8-13$ pixels (16-25 \kms), depending on wavelength.  A
scattered light correction was made by adjusting the flux at the bottom of each
neighboring saturated \HI\ line to zero.  To account for small errors in the
wavelength calibration, wavelength shifts between spectral windows were
permitted in the fit.  The fits for each \DI\ line are shown in
Figure~\ref{fig_owens}.  Displayed for each line are the observed data, the
continuum, and the fit after convolution with the instrumental LSF.

Various tests were done to investigate the magnitude of possible systematic
uncertainties associated with the profile fitting procedure.  These are
described in detail in Hebrard et al.  (2001), and include modelling the LSF
with two Gaussians, fitting the continua with polynomials up to order 14, and
varying the zero flux level from zero to twice the level at the base of each
neighboring \HI\ line.  The results of these tests helped determine the
magnitude of the \DI\ column density error.  The column density computed using
the profile fitting analysis is log $N$(\DI)$_{pf}$=$15.52\pm0.10$ ($2\sigma$).

We take the weighted mean of the results from the curve of growth and profile
fitting procedures to arrive at our best estimate of the \DI\ column density,
log $N$(\DI)=$15.47\pm0.06$ ($2\sigma$).  To be conservative, we have not
reduced the error below that determined from the COG result because the error is
dominated by systematic effects and cannot be combined in the usual statistical
fashion.

\subsection{The \OI\ Analysis}

The \OI\ column density is derived using only a COG analysis.  Most of the \OI\
lines are saturated or nearly so, which makes a profile fitting analysis
difficult, especially when there are significant uncertainties in the LSF and
velocity structure of the absorption profiles, as is the case in this study.  In
contrast to the \DI\ analysis, no stellar model was removed prior to measuring
the \OI\ lines since the stellar profile is only slowly varying in the vicinity
of these lines.  The local continuum was fit using the same procedure described
above for measuring the equivalent widths.  No strong \htwo\ lines are blended
with any of the \OI\ lines used in the analysis, so no correction for \htwo\ was
required.  Small residual flux contributed by scattered light or a broad LSF has
a larger effect on the estimated equivalent widths of the more saturated \OI\
lines than the \DI\ lines.  Therefore, for each \OI\ line a correction was made
by measuring the flux at the bottom of a neighboring saturated \HI\ line, the
magnitude of which was typically $\sim$2\% of the local continuum flux.
Including this correction had a negligible effect on the calculated column
density, but slightly increased the estimated error.

\begin{figure}
\plotone{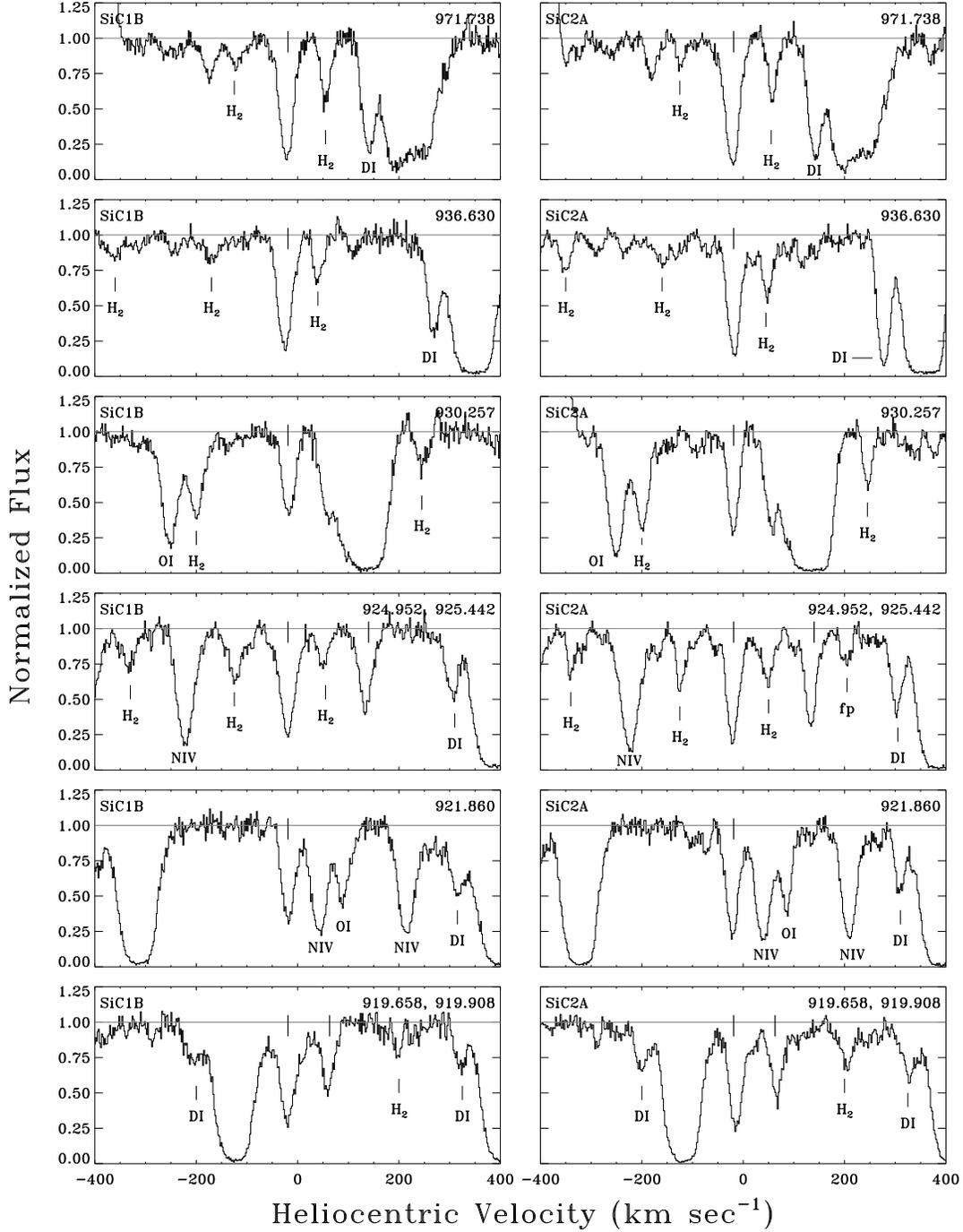}
\caption{Continuum normalized \OI\ absorption profiles.  The positions of the
\OI\ lines used in the column density analysis are indicated by the vertical
tick marks above the spectra.  These are the only \OI\ lines for which a
continuum level could be established with reasonable confidence.  The spectra
from segment SiC1B are in the left panels, and those from segment SiC2A are in
the right panels.
\label{fig_OI_cont}}
\end{figure}

The analysis was done in the same way as for the \DI\ lines, again assuming a
single-component Gaussian curve of growth.  The statistical and systematic
errors were calculated and combined in the same way as for \DI.  However, an
additional error term resulting from a 2\% uncertainty in the zero flux level
(Sembach \& Savage 1992) was combined in quadrature with this to determine the
final error.  Our best fit COG estimate gives log $N$(OI)=$16.73\pm0.10$
($2\sigma$) and $b$=$6.58\pm0.38$ \kms ($2\sigma$).  The continuum normalized
\OI\ profiles are shown in Figure~\ref{fig_OI_cont}, and the measured equivalent
widths are given in Table 5.  The COG is displayed in Figure~\ref{fig_cog}.

\section{$N$(\HI) Toward Feige 110}

In the direction of Feige 110, the \lya\ line provides the best constraint on
the total \HI\ column density because its radiation damping wings are very
strong.  \lya\ is not accessible to \fuse.  While the remaining Lyman series
lines recorded in the \fuse\ spectrum are strongly saturated, they are not
sufficiently strong to show well-developed damping wings.  We have examined the
range of $b-$value/column density combinations allowed by the higher Lyman
series lines in the \fuse\ spectrum, and we find that these lines do not
constrain $N$(\HI) with sufficient precision to provide an interesting D/H
measurement.  At the time of this writing no suitable {\it HST} spectra of the
Feige 110 \lya\ line had been obtained.  Consequently, we have used the only
high-dispersion echelle observation of this star obtained with the short
wavelength camera on \iue\ (exposure ID SWP15270) to estimate $N$(\HI) toward
Feige 110.  This \iue\ mode provides a resolution of $\sim$25 \kms\ FWHM and
covers the 1150-1950 \AA\ range.

\subsection{{\it IUE} Data}

The high-dispersion \iue\ observation was obtained on 1981 October 15 with an
exposure time of 15.3 ksec.  We reduced the data using the IUEDR processing
(Giddings \& Rees 1989).  This reduction properly traces the echelle orders and
corrects for scattered light.  Note that both the IUESIPS and NEWSIPS
reductions, which are available from the STScI MAST archive, have significant
difficulties which make them less suitable for this analysis.  The IUESIPS
reduction was performed with the an early version of IUESIPS software, and was
never reprocessed with the later, improved version.  It apparently does not
trace the orders properly, and therefore some of the signal is improperly
excluded, giving rise to undulations in the extracted spectrum.  In the NEWSIPS
reduction the background subtraction is poor.  In the fully saturated \lya\
core, which should be centered on zero flux, the NEWSIPS spectrum shows a
significant residual flux equal to $\sim$16$- $19\% of the continuum flux.
Furthermore, the core is not flat; the residual flux is higher on the red side
than on the blue side.  The IUEDR reduction does not suffer from any of these
difficulties.

\subsection{Analysis Method}

To measure the total interstellar \HI\ column density, we used the method of
Jenkins (1971); see also Jenkins et al.  (1999) and Sonneborn et al.  (2000).  In
brief, we constrained the \HI\ column density using the Lorentzian wings of the
\lya\ profile, which have optical depth $\tau$ at wavelength $\lambda$ given by
$\tau (\lambda ) = N$(\HI)$\sigma (\lambda ) = 4.26 \times 10^{-20}
N$(\HI)$(\lambda - \lambda _{0})^{-2}$.  Here $\lambda _{0}$ is the centroid of
the interstellar \HI\ absorption in this direction, which was determined from
the \ion{N}{1} triplet at 1200 \AA .  We estimated the value of $N$(\HI) that
provides the best fit to the \lya\ profile by minimizing $\chi ^{2}$ using
Powell's method with five free parameters:  (1) $N$(\HI), (2)-(4) three
coefficients that fit a second-order polynomial to the continuum (with a model
stellar \lya\ line superimposed, see below), and (5) a correction for the flux
zero level.  We then increased (or decreased) $N$(\HI) while allowing the other
free parameters to vary in order to set upper and lower confidence limits based
on the ensuing changes in $\chi ^{2}$.

\begin{figure}
\plotone{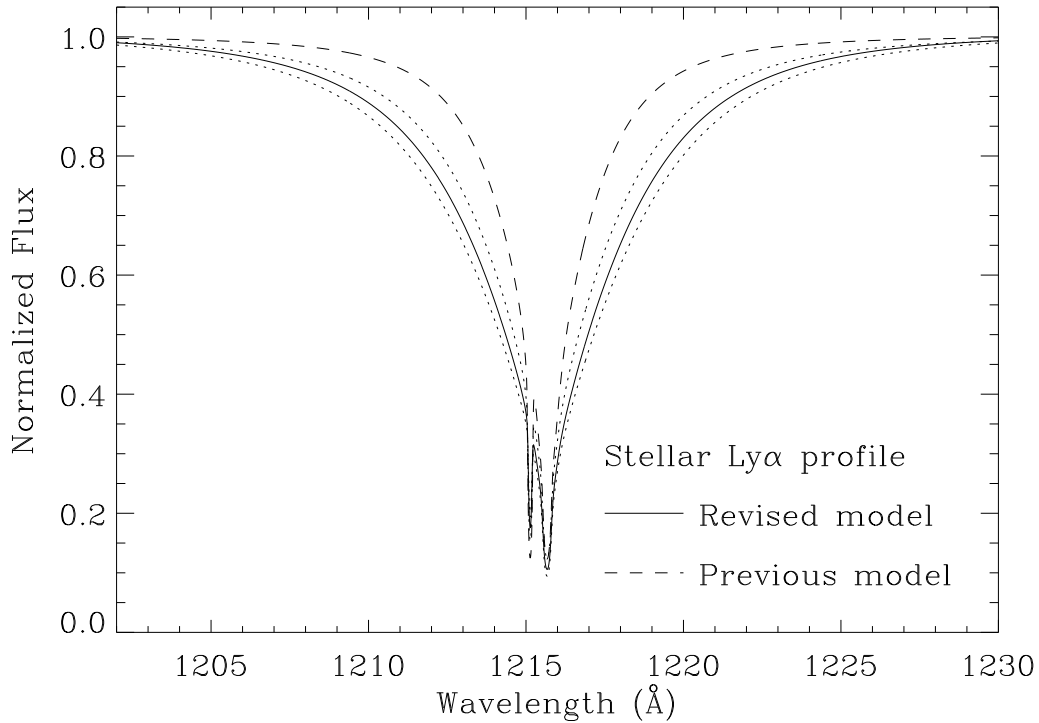}
\caption{Synthetic stellar spectra of Feige 110 in the vicinity of the stellar 
\lya\ line, normalized to the adjacent continuum. The \lya\ profile 
resulting from our preferred parameters for the stellar atmosphere (see \S 3) 
is plotted with a solid line, and the dotted lines show the highest and 
lowest profiles found in a grid of 27 models spanning the expected range 
of $T_{\rm eff}$, log $g$, and He/H for this star. The revision of the 
stellar model has a significant impact on the stellar \lya\ profile. To show 
this, the dashed line indicates the profile which would result from the 
previously published analysis (see \S 3).
\label{stlya}}
\end{figure}

In addition to the \lya\ absorption profile due to interstellar \HI\ , a
subdwarf star like Feige 110 will have a substantial stellar \lya\ absorption
line with broad, Lorentzian wings as well.  Neglect of this stellar line when
setting the continuum for fitting the interstellar \lya\ will lead to a
substantial systematic overestimate of $N$(\HI).  We have used the stellar
atmosphere models discussed in \S 3 to account for the stellar \lya\ line.
Figure~\ref{stlya} shows with a solid line the continuum- normalized stellar
\lya\ + He absorption profile predicted for the most likely combination of
$T_{\rm eff}$, log $g$, and He/H for this star.  We superimposed the normalized
stellar profile on a second-order polynomial to allow for possible instrumental
calibration problems (see above).  The offset of the stellar line centroid with
respect to the interstellar lines was determined by comparing the velocities of
several stellar and interstellar lines in the \fuse\ spectrum, and the
uncertainty in the determined offset has a negligible impact on the derived
$N$(\HI).  The dotted lines in Figure~\ref{stlya} indicate the shallowest and
deepest stellar profiles derived from a grid of models covering the expected
range of atmospheric parameters for Feige 110.  We use these ``extreme'' stellar
profiles to set upper and lower confidence limits on $N$(\HI) in the following
section.  For purposes of illustration, the dashed line shows the stellar \lya\
line predicted for the atmospheric parameters derived for Feige 110 by Heber et
al.  (1984); the impact of the stellar model revision on the derived
interstellar $N$(\ion{H}{1}) is discussed below.

\subsection{$N$(\HI) Results}

\begin{figure}
\plotone{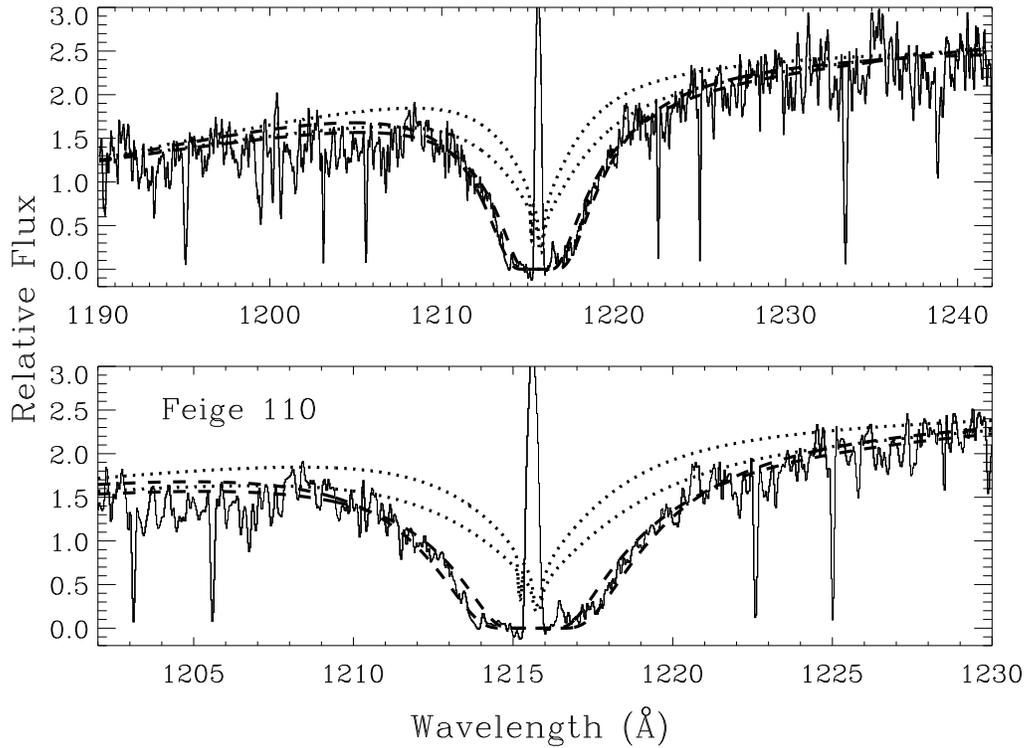}
\caption{High-dispersion {\it IUE} observation of Feige 110 (histogram) with the
fitted profiles (dashed lines) for the upper and lower bounds on $N$(\ion{H}{1})
at the 2$\sigma$ confidence level.  The continua corresponding to the upper and
lower bounds are shown with dotted lines; the upper bound uses the higher
continuum placement.  The upper and lower panels plot the same data and fits on
different scales to enable the reader to inspect the details in the \lya\
profile as well as the continuum fit well away from the interstellar \ion{H}{1}
line.  The large, narrow spikes which dip down to nearly zero flux are artifacts
due to the joining of echelle orders in the \iue\ spectrum or reseau marks on
the face plate of the \iue\ camera.
\label{hifit}}
\end{figure}

Figure~\ref{hifit} shows the \iue\ spectrum of Feige 110; the same data are
plotted in both panels but the upper panel shows a broader wavelength range to
enable the reader to inspect the fit to continuum away from the strong \lya\
absorption.  Overplotted on the spectrum with dashed lines are the fits that
provide 2$\sigma$ upper and lower limits on the interstellar $N$(\HI).  The
continua adopted for these 2$\sigma$ limits are shown with dotted lines (the
higher continuum corresponds to the upper limit).  As noted above, in addition
to the differing overall continuum level, the deepest model stellar \lya\ line
is assumed for the lower limit on $N$(\HI), and the shallowest stellar \lya\
model is assumed for the upper limit.  From these fits we derive log $N$(\HI) =
20.14$^{+0.13}_{-0.20}$ (2$\sigma$).

We note that the revision of the stellar atmosphere parameters discussed in \S 3
has a substantial impact on the derived interstellar $N$(\HI).  If we use the
older, Heber et al.  (1984) atmospheric parameters to synthesize the stellar
\lya\ line, we obtain the stellar profile shown with a dashed line in
Figure~\ref{stlya}.  With this stellar profile, we obtain an \HI\ column density
which is $\sim 0.05-0.2$ dex higher than that derived with our preferred stellar
atmosphere parameters (see Table 2), where this range of values depends on the
choice of continuum placement.  Evidently, with this type of star, uncertainties
in the stellar \lya\ line can introduce significant systematic errors into
interstellar $N$(\HI) estimates.  This also demonstrates the need for high
quality optical data to tightly constrain the stellar model parameters.

There are, of course, many potential sources of error in estimating $N$(\HI)
based only on an \iue\ \lya\ spectrum.  To improve the estimate we attempted to
measure $N$(\HI) by fitting the \lyb\ and \lyg\ lines in the \fuse\ data.
However, several difficulties prevent these from further constraining $N$(\HI).
For example, the stellar model does not include all metal lines actually present
in the stellar spectrum (see \S3 and Fig.~\ref{fig_spec_f110}).  The deviation
of the model from the observed spectrum due to this line blanketing in many
narrow spectral windows means that the effective continuum emerging from the
star is overestimated by an unknown amount.  Although this problem exists for
all Ly lines, it is much less serious for \lya\ because in this case the
interstellar Lorentzian wings are stronger and much broader than the stellar
Lorentzian profile.  The opposite is true for \lyb\ and all higher Ly series
lines.  Thus, the continuum around \lyb\ is much more steeply sloping, and much
harder to fit properly, and this is even worse around \lyg.

Complicating the issue of proper continuum placement are the effects of
extraneous interstellar lines in the vicinity of the \HI\ and \DI\ lines.
Problematic species include \OI\ and \htwo, neither of which affect the \lya\
profile.

Because of these problems the analysis of the \lyb\ and \lyg\ lines did not
allow an improved estimate of $N$(\HI) and we have relied exclusively on the
\iue\ \lya\ data.  As discussed above, the systematic error associated with this
is due primarily to unrecognized, narrow stellar lines.  We have included the
effects of known stellar lines in the continuum regions used to estimate
$\sigma$ for the $\chi ^{2}$ calculation, but nevertheless future higher
resolution and signal-to-noise observations that will allow us to mask out the
strong stellar lines might yield a different result.  For this reason, it would
be highly valuable to re-observe Feige 110 with the Space Telescope Imaging
Spectrograph on board {\it HST}.  Sonneborn et al.  (2001) have shown, using
high-resolution STIS observations of BD+28$^{\circ}$4211, that such an
observation can yield a very precise measurement of $N$(\HI), even for a
subdwarf like Feige 110.

\section{Discussion}

We have measured the column densities of \DI, \OI, and \HI\ toward Feige 110,
the longest sight line with the highest \HI\ column density of those reported in
the initial set of deuterium studies with \fuse.  Our results are summarized in
Table 6.

The \DI\ column density was measured with two independent techniques, which give
nearly the same result, despite being subject to a much different set of
systematic errors.  This agreement reflects the fact that there are many weak,
unsaturated \DI\ lines available for analysis, and shows the benefits of
observing the high-order Lyman lines available with \fuse.  The situation for
\OI\ is not as favorable, since most of the available lines are saturated or
nearly so.  This is the reason for the larger uncertainty associated with this
column density measurement.  Nevertheless, we are able to draw useful
conclusions about the ratios of \DI, \OI, and \HI.

Using our best column density estimates we find D/H = $(2.14 \pm 0.82) \times
10^{-5} (2\sigma).$ This deviates from some, but not all, values along sight
lines both within the Local Bubble and beyond (Moos et al.  2001; Linsky 1998;
Jenkins et al.  1999; Sonneborn et al.  2000).  This suggests that the D/H ratio
toward Feige 110 differs from that along other sight lines, both longer and
shorter.  This important conclusion is discussed more fully by Moos et al.
(2001), who show that there is no evidence of variability within the local
bubble, at distances less than $\sim$100 pc, but there is significant
variability at larger distances.  This is discussed within the context of mixing
of interstellar material driven by supernova explosions.  For a uniform
supernova explosion rate, material in the region may be well mixed over scales
of $\sim$100 pc.  But if supernova events are episodic, such as those that are
triggered in OB associations, there may be insufficient time to mix the material
to achieve the uniformity observed in the current D/H studies.  Other effects,
such as the gas temperature, the volume density, and the magnetic field strength
in the ISM, must be included in the models to accurately characterize the mixing
time and length scales.  The measurements discussed here and by Moos et al.  may
constrain these models.

Our result O/H = $(3.89 \pm 1.67) \times 10^{-4} (2\sigma)$, while not very
precise, is in agreement with previous measurements of O/H.  Meyer, Jura, \&
Cardelli (1998) found remarkable uniformity, O/H = $(3.43 \pm 0.77) \times
10^{-4}$ for 13 sight lines ranging in distance from 130 to 1500 pc, where the
error is the $2\sigma$ dispersion in the measured values.  (We have increased
their published ratio by about 7.5\% to account for a change in the \OI\ 1356
oscillator strength recommended by Welty et al.  (1999).)  Moos et al.  (2001)
find no evidence of O/H variability in the sight lines studied with \fuse.

A very sensitive test of the idea that the deuterium abundance is decreasing
with time due to stellar astration is the spatial distribution of the D/O ratio.
The ionization potential of neutral H and D (13.598 eV) and O (13.618 eV) are
nearly identical.  Charge exchange reactions between H, D, and O keep the
ionization fractions coupled such that \DI /\OI\ is a good approximation of D/O.
We find D/O = $(5.50 \err{1.64}{1.33}) \times 10^{-2}$.  Moos et al.  (2001)
find no statistically significant spatial variation in this quantity, although
the scatter about the mean value decreases if only the targets within the Local
Bubble are considered.  Furthermore, they find no evidence of anti-correlation
between D/O and O/H.  Although the metallicity range of these local ISM studies
is small, this may indicate that some subtle nucleosynthetic processes are
affecting the D/O ratio.  One possibility discussed by Moos et al.  is that
there are stellar processes which destroy weakly bound deuterium nuclei without
creating significant quantities of oxygen, and stellar winds mix this deuterium
depleted material into the ISM.  Additional measurements using \fuse, some of
which probe sight lines to more distant targets and are currently under study,
will shed more light on this interesting possibility.

\acknowledgements

This work is based on data obtained for the Guaranteed Time Team by the
NASA-CNES-CSA \fuse\ mission operated by the Johns Hopkins University.
Financial support to U.  S.  participants has been provided by NASA contract
NAS5-32985.  French participants are supported by CNES.  This work has used the
profile fitting procedure {\tt Owens.f} developed by M.  Lemoine and the \fuse\
French Team.  We thank Rex Saffer for sharing his optical spectrum of Feige 110.


\begin{deluxetable}{ccccc}
\tablenum{1}
\tablecolumns{5}
\tablewidth{0pc}
\tablecaption{Feige 110 Observation Log}
\tablehead{
\colhead{Instrument} &
\colhead{Date} & \colhead{Program ID} &
\colhead{Exposures} &
\colhead{Exposure Time}
}
\startdata

\fuse & 22 June 2000 & M1080801 & 001 - 008 & 6228 \\
\fuse & 30 June 2000 & P1044301 & 019 - 066\tablenotemark{a} & 21779 \\
\iue  & 15 October 1981 & \nodata & SWP 15270 & 15300 \\
\enddata
\tablenotetext{a}{Exposures 001-018 missing due to an acquisition problem
with the previous observation.}
\end{deluxetable}

\begin{deluxetable}{lcc}
\tablenum{2}
\tablecolumns{8}
\tablewidth{0pt}
\tablecaption{Target and Sight Line Parameters for Feige~110}
\tablehead{
\colhead{Quantity} &
\colhead{Value} &
\colhead{Reference}
}
\startdata
Spectral Type & sdOB & 1 \\
$(l,b)$ & $(74\fdg09,-59\fdg07)$ & 2 \\
$d$\tablenotemark{a} \ (pc)& $179 \err{265}{67}$ & 2 \\
$z$ (pc) & $-154 \err{57}{227}$ & 2 \\
$V$ & 11.86 & 3 \\
$U-B$ & $-1.19$ & 3 \\
$B-V$ & $-0.33$ & 3\\
$T_{\rm eff}$\ (K)& $42300 \pm 1000$ & 4 \\
$\log g$\ (cm s$^{-2}$) & $5.95 \pm 0.15$& 4 \\
$\log ({\rm He / H})$ & $-1.95 \pm 0.15$ & 4 \\
log $N$(\HI) & $20.18^{+0.14}_{-0.21}$ & 4 \\
$<n_H>$ & $0.27~cm^{-3}$ & 4 \\
\enddata
\tablenotetext{a}{Derived from trigonometric parallax}
\tablerefs{(1) Heber et al.\ 1984; (2)\ Perryman et al.\ 1997;
(3)\ Kidder et al.\ 1991; (4)\ this study.}
\end{deluxetable}

\begin{deluxetable}{lcccccc}
\tablenum{3}
\tablecolumns{7}
\tablewidth{0pc}
\tablecaption{\htwo\ Equivalent Widths}
\tablehead{
\colhead{} & \colhead{} & \colhead{} &
\multicolumn{4}{c}{$W_\lambda$ [m\AA]\tablenotemark{b}} \\
\cline{4-7}
\colhead{Line Identification} & \colhead{$\lambda$(\AA)} &
\colhead{$\log \lambda f$\tablenotemark{a}} & \colhead{SiC1B} & \colhead{SiC2A} &
\colhead{LiF1A} & \colhead{LiF2B}
}
\startdata
W (3-0) R(2) & 947.111 & 1.101 & $12.6\pm3.4$ & $10.4\pm1.2$ & \nodata & \nodata \\
W (3-0) Q(3) & 950.397 & 1.417 & $14.2\pm2.8$ & $14.4\pm1.9$ & \nodata & \nodata \\
W (2-0) Q(4) & 971.387 & 1.533 & $16.2\pm3.6$ & $12.0\pm2.1$ & \nodata & \nodata \\
L (6-0) P(3) & 1031.191 & 1.059 & \nodata & \nodata & $48.7\pm4.1$ & $45.6\pm2.3$ \\
\enddata
\tablenotetext{a}{Oscillator strengths from Abgrall et al. 1993a,b.}
\tablenotetext{b}{Measured equivalent widths and $1\sigma$ 
 uncertainties (in m\AA).}
\end{deluxetable}

\begin{deluxetable}{cccc}
\tablenum{4}
\tablecolumns{4}
\tablewidth{0pc}
\tablecaption{\DI\ Equivalent Widths}
\tablehead{
\colhead{} & \colhead{} & 
\multicolumn{2}{c}{$W_\lambda$ [m\AA]\tablenotemark{b}} \\
\cline{3-4}
\colhead{$\lambda$(\AA)} & \colhead{$\log \lambda f$\tablenotemark{a}} & 
\colhead{SiC1B} &
\colhead{SiC2A}
}
\startdata
919.101 & 0.0425 & $22.1\pm2.8$ & $25.3\pm2.5$ \\
920.712 & 0.171 & $23.3\pm2.5$ & $30.0\pm3.6$ \\
922.899 & 0.312 & $27.4\pm2.6$ & $30.5\pm2.4$ \\
925.973 & 0.469 & $36.1\pm2.2$ & $42.4\pm2.2$ \\
937.548 & 0.864 & $56.4\pm3.5$ & \nodata \\
949.484 & 1.121 & $70.5\pm6.1$ & $69.3\pm2.3$ \\
\enddata
\tablenotetext{a}{Oscillator strengths from Morton 1991.}
\tablenotetext{b}{Measured equivalent widths and $1\sigma$ 
 uncertainties (in m\AA).}
\end{deluxetable}

\begin{deluxetable}{cccc}
\tablenum{5}
\tablecolumns{4}
\tablewidth{0pc}
\tablecaption{\OI\ Equivalent Widths}
\tablehead{
\colhead{} & \colhead{} & 
\multicolumn{2}{c}{$W_\lambda$ [m\AA]\tablenotemark{b}} \\
\cline{3-4}
\colhead{$\lambda$(\AA)} & \colhead{$\log \lambda f$\tablenotemark{a}} & 
\colhead{SiC1B} &
\colhead{SiC2A}
}
\startdata
919.658 & -0.060 & $74.9\pm6.5$ & $71.0\pm11.8$ \\
919.917 & -0.786 & $40.7\pm6.2$ & $43.1\pm8.3$ \\
921.857 & 0.041 & $66.4\pm2.8$ & $74.0\pm5.3$ \\
924.950 & 0.155 & $81.9\pm5.7$ & $71.6\pm4.4$ \\
925.442 & -0.485 & $55.6\pm2.8$ & $55.0\pm3.7$ \\
930.257 & -0.301 & $56.2\pm3.7$ & $57.8\pm3.5$ \\
936.630 & 0.534 & $86.2\pm2.9$ & $83.9\pm7.0$ \\
971.738 & 1.123 & $100.2\pm7.7$ & $101.4\pm3.5$ \\
\enddata
\tablenotetext{a}{Oscillator strengths from Morton 1991 and Morton (private
communication).}
\tablenotetext{b}{Measured equivalent widths and $1\sigma$ 
 uncertainties (in m\AA).}
\end{deluxetable}

\begin{deluxetable}{lc}
\tablenum{6}
\tablecolumns{2}
\tablewidth{0pc}
\tablecaption{Summary of Results of This Study}
\tablehead{
\colhead{Quantity} &
\colhead{Value ($2\sigma$ errors)}
}
\startdata
log $N$(\DI) & $15.47\pm0.06$ \\
log $N$(\OI) & $16.73\pm0.10$ \\
log $N$(\HI) & $20.14^{+0.13}_{-0.20}$ \\
D/H & $(2.14 \pm 0.82) \times 10^{-5}$ \\
O/H & $(3.89 \pm 1.67) \times 10^{-4}$ \\
D/O & $(5.50 \err{1.64}{1.33}) \times 10^{-2}$ \\
\enddata
\end{deluxetable}


\begin{references}

\reference{} Abgrall, H., Roueff, E., Launay, F., Roncin, J.Y., \& Subtil, J.L., 1993a,
   \aaps, 101, 273

\reference{} Abgrall, H., Roueff, E., Launay, F., Roncin, J.Y., \& Subtil, J.L., 1993b,
   \aaps, 101, 323

\reference{} Bluhm, H., Marggraf, O., de Boer, K.  S., Richter, P., \& Heber,
U. 1999, \aap, 352, 287.

\reference{} Boesgard, A. \& Steigman, G. 1985, \araa, 23, 319

\reference{} Burles, S. \& Tytler, D. 1998, \apj, 499, 699

\reference{} Caloi, V.\ 1989, \aap, 221, 27

\reference{} Dorman, B., Rood, R.\ T., \& O'Connell, R.\ W.\ 1993,
  \apj, 419, 596
  
\reference{} Epstein, R. I., Lattimer, J. M., \& Schramm, D. N., 1976, \nat, 
   263, 198.

\reference{} Feige, J.\ 1958, \apj, 128,267

\reference{} Giddings, J. R., \& Rees, P. C. T., 1989, SERC Starlink User Note 37

\reference{} G\"{o}lz, M. et al. 1998, in Proc. IAU Colloq. 166, The Local
Bubble and Beyond, ed. D. Breitschwerdt, M. J. Freyberg, \& J. Trumper 
(Berlin: Springer), 75

\reference{} Greenstein, J.\ L., \& Sargent, A.\ I.\ 1974, \apjs, 28, 157

\reference{} Heber, U., Hamann, W.-R., Hunger, K., Kudritzki, R.\ P.,
  Simon, K.\ P., \& M\'endez, R.\ H.\ 1984, \aap, 136, 331

\reference{} H\'ebrard, G. et al. 2001, \apj, submitted

\reference{} Hubeny, I, \& Lanz, T.\ 1995, \apj, 439, 875

\reference{} Jenkins, E. B. 1971, \apj, 169, 25

\reference{} Jenkins, E. B. 1986, \apj, 304, 739.

\reference{} Jenkins, E. B., Tripp, T. M., Wo\'{z}niak, P. R., Sofia, U. J., 
\& Sonneborn, G. 1999, ApJ, 520, 182

\reference{} Kidder, K.\ M., Holberg, J.\ B., \& Mason, P.\ A.\ 1991,
  \aj, 101, 579

\reference{} Kirkman, D., Tytler, D., Burles, S., Lubin, D., \& O'Meara, J. M.,
2000, \apj, 529, 655

\reference{} Kruk, J.W. et al. 2001, \apj, submitted

\reference{} Kurucz, R., \& Bell, B.\ 1995, Atomic Line Data
  Kurucz CD-ROM No. 23, (Cambridge, Mass.: Smithsonian Astrophysical 
  Observatory)
  
\reference{} Lallement, R. \& Bertin, P. 1992, \aap, 266, 479

\reference{} Lehner, N. et al. 2001, \apj, submitted

\reference{} Lemoine, M. et al. 2001, \apj, submitted

\reference{} Linsky, J. L. et al. 1995, \apj, 451, 335

\reference{} Linsky, J. L. 1998, Space Sci. Rev. 84, 285

\reference{} Meyer, D. M., Jura, M., \& Cardelli, J. A., \apj, 493, 222.

\reference{} Morton, D.C. 1991, \apjs, 77, 119

\reference{} Moos, H. W. et al. 2000, \apj, 538, L1

\reference{} Moos, H. W. et al. 2001, \apj, submitted

\reference{} Perryman, M.\ A.\ C., et al.\ 1997, \aap, 323, L49

\reference{} Reeves, H., Audouze, J., Fowler, W., \& Schramm, D., N. 
\apj, 179, 909

\reference{} Rogerson, J. B. \& York, D. G. 1973, \apj, 186, L95

\reference{} Sahnow, D. J. et al. 2000, \apj, 538, L7

\reference{} Sahu, M. S. et al. 1999, \apj, 523, L159.

\reference{} Sembach, K. R. \& Savage, B. D. 1992, \apjs, 83, 147

\reference{} Sfeir, D. M., Lallement, R., Crifo, F., \&
Welsh, B. Y.  1999, \aap, 346, 785

\reference{} Sonneborn, G., Tripp, T.  M., Ferlet, R., Jenkins, E.  B., Sofia, U.
J., Vidal-Madjar, A., \& Wo\'{z}niak, P.  R.  2000, ApJ, 545, 277 \

\reference{} Sonneborn, G. et al. 2001, \apj, submitted

\reference{} Spitzer, L. 1978, Physical Processes in the Interstellar Medium
(New York: John Wiley)

\reference{} Tenorio-Tagle 1996, \aj, 111, 1641

\reference{} Tosi, M. et al. 1998, \apj, 498, 226

\reference{} Vidal-Madjar, A. \& Gry, C. 1984, \aap, 138, 285.

\reference{} Vidal-Madjar, A. et al. 1998, \aap, 338, 694.

\reference{} Welty, D.E. et al. 1999, \apjs, 124, 465

\reference{} Wood, B.E. et al. 2001, \apj, submitted

\end{references}
\end{document}